\documentclass[a4paper]{jpconf}
\usepackage{graphicx}

\newcommand{\lssim}{\mbox{\raisebox{-.9ex}{~$\stackrel{\mbox{$<$}}{\sim}$~}}}
\newcommand{\gsim}{\mbox{\raisebox{-.9ex}{~$\stackrel{\mbox{$>$}}{\sim}$~}}}
\newcommand{\calp}{{\cal P}}
\newcommand{\fnl}{f_{\rm NL}}

\begin{document}
\title{
The quantum origin of cosmic structure: Theory and observations}

\author{Konstantinos Dimopoulos}

\address{
Physics Department, Lancaster University, Lancaster LA1~4YB, UK}

\ead{k.dimopoulos1@lancaster.ac.uk}

\begin{abstract}
The particle production process is reviewed, through which cosmic inflation 
can produce a scale invariant superhorizon spectrum of perturbations of suitable
fields starting from their quantum fluctuations. Afterwards, in the context of 
the inflationary paradigm, a number of mechanisms (e.g. curvaton, inhomogeneous
reheating etc.) through which such perturbations can source the curvature 
perturbation in the Universe and explain the formation of structures such as 
galaxies are briefly described. Finally, the possibility that cosmic vector 
fields also contribute to the curvature perturbation (e.g. through the vector 
curvaton mechanism) is considered and its distinct observational signatures are 
discussed, such as correlated statistical anisotropy in the spectrum and 
bispectrum of the curvature perturbation.
\end{abstract}

\section{Introduction}
The standard model of cosmology at present is comprised by Hot Big Bang 
Cosmology and Cosmic Inflation. The cosmology of the Hot Big Bang accounts
for the Hubble expansion of the Universe, the observed Cosmic Microwave 
Background (CMB) radiation, the primordial abundance of light elements
(formed in the Early Universe through the process of Big Bang Nucleosynthesis)
and the age of the Universe, which agrees well with the astrophysical estimates
of the ages of the oldest globular clusters. In turn inflation overcomes or at
least ameliorates certain fine-tuning problems regarding the initial conditions
of the Hot Big Bang, namely the so-called horizon and flatness problems
\cite{guth} \cite{staro}.

What is cosmic inflation? In a nutshell, inflation is a brief period of 
superluminal expansion of space in the Early Universe. What inflation does is 
that it makes the observable Universe large and uniform. However, if the 
Universe were perfectly uniform there would be no structures like galaxies
or galactic clusters, no stars with planets orbiting around them, no ... us.
It is imperative, therefore, that there is a deviation from perfect uniformity,
which can give rise to these structures. Indeed, we need a Primordial Density
Perturbation (PDP) for structure formation to occur. It so happens that 
evidence for such a PDP exists as the latter reflects itself on the CMB through
the so-called Sachs-Wolfe effect \cite{sachs}
which states that CMB light is redshifted when
crossing growing overdensities. This effect directly connects the fractional
amplitude of the PDP with the fractional perturbation of the temperature of the
CMB:
\begin{equation}
\left.\frac{\delta T}{T}\right|_{\rm CMB}=\frac12
\left.\frac{\delta\rho}{\rho}\right|_H\approx 10^{-5}.
\end{equation}
Even though the PDP appears to be very small, numerical simulations of 
structure formation have shown that it is enough to account for the observed 
structure in the Universe. What is the origin of this PDP? Well, it turns out 
that this too can be accounted for by Cosmic Inflation.

\section{Particle production during cosmic inflation}\label{pp}
To have an idea of how Cosmic Inflation is achieved consider the so-called
Friedman equation, which is the temporal component of the Einstein equations
for a homogeneous and isotropic spacetime.
In flat space this equation reeds
\mbox{$H^2=\frac13\rho/m_P^2$}, where \mbox{$m_P^2=(8\pi G)^{-1}$} is the 
reduced Planck mass\footnote{$G$ is Newton's gravitational constant and we 
consider natural units where \mbox{$c=\hbar=k_B=1$}.} 
and \mbox{$H\equiv\dot a/a$} is the Hubble parameter corresponding to the rate 
of the expansion of the Universe, with $a=a(t)$ being the scale factor of the 
Universe, parameterising the Universe expansion, and the dot denotes derivative
with respect to the cosmic time $t$. 
Suppose now that, at some period in its early history, 
the Universe was dominated by an effective cosmological constant 
$\Lambda_{\rm eff}$. Then, \mbox{$\rho\simeq m_P^2\Lambda_{\rm eff}=$ 
constant}, which means that the
Hubble parameter \mbox{$H=\Lambda_{\rm eff}/3$} is constant and, therefore, 
\mbox{$a\propto e^{Ht}$}, i.e. space expands exponentially in time. Thus, 
inflation occurs when the Universe is dominated by an effective vacuum density.
When inflation ends this density has to be transferred in the density of the
thermal bath of the Hot Big Bang. This, in effect, amounts to a change of 
vacuum\footnote{from the false vacuum of inflation, corresponding to 
$\Lambda_{\rm eff}$, to the true vacuum of the present, with 
$\Lambda\simeq 0$.}. Therefore, vacuum states during inflation are not 
necessarily vacuum states after inflation, but instead they can become 
populated. 

This process is called Particle Production and it is similar to the
production of particles (in the form of Hawking radiation) on the event horizon
of Black Holes \cite{hawk}. 
Indeed, the cosmological horizon during inflation is an event 
horizon (of size $\sim H^{-1}$)
and can be viewed as an ``inverted'' (i.e. inside-out) black hole in 
the sense that nothing can escape being ``sucked out'' by the superluminal 
expansion. Virtual particle pairs, corresponding to quantum fluctuations, are
broken up by the expansion and are pulled away to superhorizon distances, where
they can no more find eachother and annihilate, becoming thereby real 
particles, giving rise to classical perturbations of the corresponding fields
\cite{contp}.
The amplitude of these perturbations is determined by the Hawking temperature
$\delta\phi\sim T_H$, which for de Sitter space is \mbox{$T_H=H/2\pi$} 
\cite{gibb}. Let us
review now the particle production process in a more rigorous manner.

The standard paradigm considers a real, minimally coupled, scalar field $\phi$
of mass $m$ with Lagrangian density
\begin{equation}
{\cal L}=\frac12\partial_\mu\phi\,\partial^\mu\phi-\frac12m^2\phi^2.
\label{Lphim}
\end{equation}
Using the above one obtains the equation of motion of the scalar field. The 
field is expected to become homogenised by the inflationary expansion as any
inhomogeneities are inflated away. In this case the equation of motion becomes
\begin{equation}
\ddot\phi+3H\dot\phi+m^2\phi=0\,,
\label{eomhom}
\end{equation}
where \mbox{$\phi=\phi(t)$}. Deviations from homogeneity are introduced
originally only from its vacuum fluctuations. To follow their evolution we 
perturb the field from its homogeneous value as 
\mbox{$\phi=\phi(t)+\delta\phi(\vec x, t)$}. Then we obtain the equation of
motion of the Fourier components of the field perturbations
\mbox{$\delta\phi_k(\vec k)\equiv\int\delta\phi(\vec x)
e^{-i\vec k\cdot\vec x}d\vec x$}. This equation reads
\begin{equation}
\left[\partial_t^2+3H\partial_t+m^2+\left(\frac{k}{a}\right)^2\right]
\delta\phi_k=0\,,
\label{eom}
\end{equation}
where \mbox{$k\equiv|\vec k|$}.

The next step is to promote the perturbations of the field to quantum 
operators defined as
\begin{equation}
\delta\hat\phi(\vec x,t)=\int\frac{d^3k}{(2\pi)^3}\left[
\hat a(\vec k)\delta\varphi_k(k,t)e^{i\vec k\cdot\vec x}+
\hat a^\dagger(\vec k)\delta\varphi^*_k(k,t)e^{-i\vec k\cdot\vec x}\right],
\end{equation}
where $\hat a(\vec k)$ and $\hat a^\dagger(\vec k)$ are creation and 
annihilation operators respectively and we consider canonical quantisation with
\mbox{$[\hat a(\vec k),\hat a^\dagger(\vec k')]=
(2\pi)^3\delta(\hat k-\hat k')$}. The mode functions $\delta\varphi_k(k,t)$
satisfy the same equation of motion as the Fourier components of the field
perturbations $\delta\phi_k(\vec k,t)$ because this equation is linear. We can
solve this equation considering that in the subhorizon limit 
(\mbox{$k/aH\rightarrow+\infty$}) the solution matches the so-called 
Bunch-Davies vacuum \cite{BD}
which corresponds to the quantum fluctuations of a free 
scalar field in Minkowski spacetime\footnote{Locally spacetime is flat. The 
subhorizon limit is well within the radius of curvature of spacetime 
during inflation so curvature can be ignored. Similarly, in this limit the
momentum of the virtual particles is much larger than their mass 
\mbox{$k/a\gg m$} so that the field can be considered effectively massless.}.
Thus, the boundary condition reads
\begin{equation}
\delta\varphi_k
\mbox{\raisebox{-1ex}{~$\stackrel{\mbox{\Large $\longrightarrow$}}{{
\mbox{\tiny $k\!/\!a\!H\!\!\rightarrow\!\!+\!\infty$}}}$~}}
\frac{e^{-ik\tau}}{\sqrt{2k}}=
\frac{e^{ik/aH}}{\sqrt{2k}},
\end{equation}
where \mbox{$\tau=-1/aH$} is the conformal time, which factors out the 
expansion of the Universe. Using the above, the solution of Eq.~(\ref{eom})
for the mode functions is
\begin{equation}
\delta\varphi_k=a^{-3/2}\sqrt{\frac{\pi}{H}}
\frac{e^{i\frac{\pi}{2}(\nu-\frac12)}}{1-e^{i2\pi\nu}}
[J_\nu(k/aH)-e^{i\pi\nu}J_{-\nu}(k/aH)],
\label{phisolu}
\end{equation}
where $J_\nu$ denotes Bessel functions of the first kind and 
\mbox{$\nu\equiv\sqrt{\frac94-(\frac{m}{H})^2}$}. To investigate particle 
production we evaluate the above solution in the superhorizon limit
(\mbox{$k/aH\rightarrow 0$}). We find
\begin{equation}
\delta\varphi_k
\mbox{\raisebox{-1ex}{~$\stackrel{\mbox{\Large $\longrightarrow$}}{{
\mbox{\tiny $k\!/\!a\!H\!\!\rightarrow\!\!0$}}}$~}}
a^{-3/2}\sqrt{\frac{\pi}{H}}
\frac{e^{i\frac{\pi}{2}(\nu-\frac32)}}{\sin(\pi\nu)}
\frac{2^{\nu-1}}{\Gamma(1-\nu)}\left(\frac{aH}{k}\right)^\nu.
\end{equation}

Using the above we can calculate the power spectrum 
\mbox{$\calp_{\delta\phi}\equiv\frac{k^3}{2\pi^2}|\delta\varphi|^2$} in the
superhorizon limit. We find
\begin{equation}
\calp_{\delta\phi}=\frac{8\pi|\Gamma(1-\nu)|^{-2}}{1-\cos(2\pi\nu)}
\left(\frac{H}{2\pi}\right)^2\left(\frac{k}{2aH}\right)^{3-2\nu}.
\label{Pphi}
\end{equation}
Thus, we see that the scale dependence of the field perturbations is only 
included in the last term of the above. Considering a light field, i.e. 
\mbox{$m\ll H\Rightarrow\nu\rightarrow\frac32$}, we see that the scale 
dependent term is eliminated and the spectrum becomes scale invariant.
Indeed, for a light field the above spectrum can be written as
\begin{equation}
\calp_{\delta\phi}=\left(\frac{H}{2\pi}\right)^2
\left(\frac{k}{2aH}\right)^{2\eta},
\label{Pdphi}
\end{equation}
where \mbox{$\eta=\frac13(\frac{m}{H})^2$} and we considered 
\mbox{$|\eta|\ll 1$}. Thus, a light minimally coupled scalar field obtains a 
scale invariant spectrum of perturbations when they exit the horizon. The 
typical value of the field perturbations is 
\mbox{$\delta\phi\approx\sqrt{\calp_{\delta\phi}}\simeq H/2\pi$}, i.e. it is
determined by the Hawking temperature as expected.

\begin{figure}
\begin{center}
\includegraphics[width=4in]{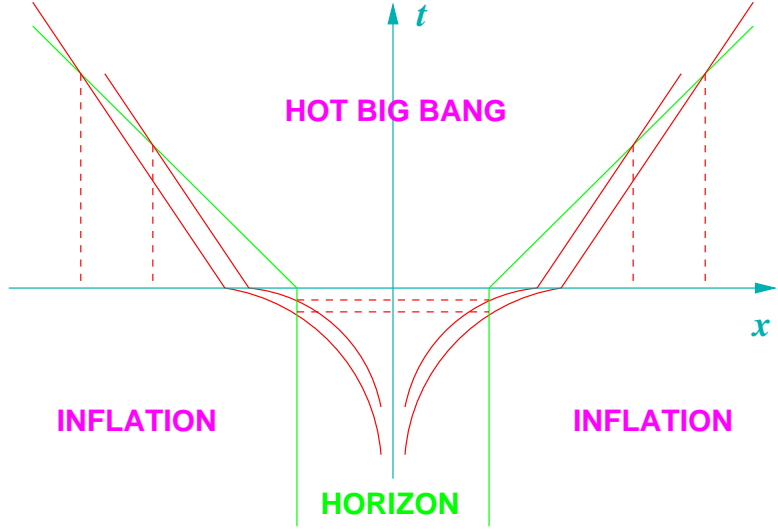}
\end{center}
\caption{\label{superhor}
Schematic log-log plot of the evolution of the perturbations of a scalar field 
(enveloped by red solid lines) which follow the Universe expansion compared 
with the cosmological horizon (solid green lines). During inflation the 
cosmological horizon is an event horizon of constant physical radius 
$\sim H^{-1}$. 
The perturbations of the scalar field start off as quantum fluctuations at 
subhorizon size, which are stretched to superhorizon scales by the 
(quasi)exponential expansion. After inflation, the size of the superhorizon 
perturbations continues to grow following the expansion of the Universe (i.e. 
proportional to $a(t)$) but the cosmological horizon (which is now a particle 
horizon) grows faster, with the speed of light. As a result, the 
perturbations reenter the horizon some time after the end of inflation. In 
the graph two different perturbations are depicted. They exit the horizon at 
different times, corresponding to the horizontal dashed lines. As a result they
are inflated to different sizes so one is much larger
that the other when they reenter the horizon. The size of the perturbations at
horizon reentry is depicted by the vertical dashed lines. However, both the 
perturbations have approximately the same amplitude, determined by the
Hawking temperature during inflation \mbox{$\delta\phi=\frac{H}{2\pi}=T_H$},
which remains approximately constant. The fact that the scalar field 
perturbations retain the same amplitude even though they attain different sizes
during their superhorizon evolution is the reason behind the scale-invariance 
of the perturbation spectrum.%
}
\end{figure}

What is the fate of these perturbations after horizon exit? The perturbations
can now be treated classically and their evolution is determined by the 
equivalent of Eq.~(\ref{eomhom})
\begin{equation}
\ddot{(\delta\phi)}+3H\dot{(\delta\phi)}+m^2(\delta\phi)=0\,.
\end{equation}
Considering that \mbox{$H\approx$ constant} the solution of the above for a 
light field is
\begin{equation}
\delta\phi=\frac{H}{2\pi}\left[e^{-\frac13(\frac{m}{H})^2H\Delta t}-
\frac19\left(\frac{m}{H}\right)^2e^{-3H\Delta t}\right]\approx\frac{H}{2\pi}.
\end{equation}
Thus, we see that the perturbations of a light scalar field freeze-out after
horizon exit. This guarantees that scale invariance is maintained since, 
regardless
of the time of horizon exit (i.e. regardless of how large the size of the 
perturbations becomes by inflation), the perturbations have the same amplitude
because \mbox{$H\approx$ constant} until the end of inflation
(see Fig.~\ref{superhor}).

How are such field perturbations related with the PDP? The PDP arises because 
of the generation of a corresponding curvature perturbation $\zeta$, as is
discussed below. If the curvature perturbation is due to the perturbations of
a light scalar field then their power spectra are proportional, i.e.
\mbox{$\calp_\zeta\propto\calp_{\delta\phi}$}. This means that they have the 
same scale dependence. The latter can be parametrised in the form of a 
power-law: \mbox{$\calp_\zeta\propto k^{n_s-1}$}, so that
\begin{equation}
n_s(k)-1\equiv\frac{d\ln\calp_\zeta}{d\ln k}\,.
\end{equation}
Assuming that $\zeta$ is due to a light scalar field one obtains
(cf. Eq.~(\ref{Pdphi}))
\begin{equation}
n_s=1+2\eta+{\cal O}(\varepsilon)\,,
\end{equation}
where \mbox{$\varepsilon\equiv-\dot H/H^2\ll 1$} quantifies the deviation from
pure de Sitter expansion which we have ignored so far. Since, during inflation
\mbox{$|\eta|\ll 1$} for a light scalar field, we find that, if the curvature 
perturbation is due to this field then \mbox{$n_s\simeq 1$} and the scale 
dependence of $\calp_\zeta$ is eliminated. Thus, the PDP in this case would be 
approximately scale-invariant. Indeed, the latest WMAP observations suggest
\cite{wmap}
\begin{equation}
n_s=0.963\pm0.012\,,
\end{equation}
which agrees with the predictions of inflation with the quantum fluctuations 
of a light scalar field as the source of the PDP. Note that the observations 
deviate from exact scale invariance (\mbox{$n_s=1$}) at 1-$\sigma$. This 
reveals some dynamics during inflation and agrees with the expectations of 
realistic inflation models.

\section{The curvature perturbation} 
In general relativity the curvature of spacetime and the energy density of its 
content are interchangeable quantities, through the Einstein equations, 
depending on the
choice of foliation of spacetime. Therefore, for the curvature perturbation it
is useful to define a quantity $\zeta$ which is independent of such foliation 
(gauge invariant). This can be written as \cite{zeta}
\begin{equation}
\zeta\equiv-\psi-H\frac{\delta\rho}{\dot\rho}\,,
\end{equation}
where the first term on the right-hand-side
is the curvature perturbation in uniform density slices of
spacetime and the second term is the density perturbation in flat slices of 
spacetime. We call $\zeta$ the (gauge invariant) curvature perturbation from 
now on. 

One can also define the power spectrum of the curvature perturbation
\mbox{$\langle\zeta^2(\vec x)\rangle\!=\!
\int_{_0}^{^\infty}\!\!d(\ln k)\calp_\zeta(k)$}
which is given by the two-point correlator as
\begin{equation}
\langle\zeta(\vec k)\zeta(\vec k')\rangle=(2\pi)^2
\delta^{(3)}(\vec k+\vec k')\frac{(2\pi)^3}{4\pi k^3}\calp_\zeta(\vec k).
\end{equation}
The latest observations of WMAP give \cite{wmap}
\begin{equation}
\sqrt{\calp_\zeta(k_0)}=(4.94\pm0.09)\times 10^{-5},
\end{equation}
where \mbox{$k_0=0.002\,$Mpc$^{-1}$} is the pivot scale. 
The corresponding density perturbation (at horizon re-entry) is given by
\begin{equation}
\left(\frac{\delta\rho}{\rho}\right)_H=
\frac25\,\zeta_{_{\rm LS}}=(1.98\pm0.04)\times 10^{-5},
\label{drobs}
\end{equation}
which, as mentioned, is the observed measurement of the PDP.\footnote{The
subscript `LS' refers to the last-scattering surface, where the CMB is 
emitted.}

Another useful quantity is the so-called bispectrum $B_\zeta$ of the curvature
perturbation, defined by the three-point correlator of $\zeta$ as follows
\begin{equation}
\langle\zeta(\vec k)\zeta(\vec k')\zeta(\vec k'')\rangle=(2\pi)^2
\delta^{(3)}(\vec k+\vec k'+\vec k'') 
B_\zeta(\vec k,\vec k',\vec k'')\,.
\end{equation}
The bispectrum can be related with the power spectrum is the following manner
\begin{equation}
B_\zeta(\vec k_1,\vec k_2,\vec k_3)=-\frac65 \fnl
[
P_\zeta(k_1)P_\zeta(k_2)+
P_\zeta(k_2)P_\zeta(k_3)+
P_\zeta(k_3)P_\zeta(k_1)
],
\end{equation}
where \mbox{$P_\zeta(k)\equiv\frac{2\pi^2}{k^3}\calp_\zeta(k)$}. 

The bispectrum is useful because it is exactly zero if the curvature 
perturbation is Gaussian, i.e. if it obeys Gaussian statistics. One expects 
that the perturbations of the 
fields which are generated during inflation from their quantum fluctuations 
are indeed Gaussian since the quantum fluctuations are random. Any non-Gaussian
signal therefore would arise by the process which translates the perturbations
of such fields, e.g. $\delta\phi$, into the curvature perturbation $\zeta$. 
This process may be highly non-linear in which case $\zeta$ would feature
non-Gaussian statistics. The level of such non-Gaussianity is quantified in the
bispectrum of $\zeta$ by the so-called non-linearity parameter $\fnl$. 
The latter also depends on the configuration of the three $\vec k$-vectors
used to determine the bispectrum.

The two most popular configurations used to determine $\fnl$ are 1) the 
equilateral configuration, where \mbox{$k_1=k_2=k_3$} and 2) the squeezed
configuration, where \mbox{$k_1=k_2\gg k_3$}. The WMAP findings for the values
of $\fnl$ in these configurations are \cite{wmap}
\begin{equation}
\fnl^{\rm eql}=26\pm140\quad{\rm and}\quad\fnl^{\rm sqz}=32\pm21\,.
\label{fnlobs}
\end{equation}
Note that there is a hint (at 1-$\sigma$) of non-Gaussianity in the squeezed
configuration. It is likely that non-Gaussianity may be detected in the near 
future by the observations of the Planck satellite, which are expected to 
improve the precision of the measurement of $\fnl$ by about an order of 
magnitude. It is important to stress here that the PDP is highly Gaussian. 
Indeed, remembering that \mbox{$\calp_\zeta(k_0)\sim 10^{-9}$} and also 
\mbox{$B_\zeta\propto\fnl\calp_\zeta^2$} we see that, even if the observational
upper bounds on $\fnl$ are saturated, non-Gaussianity in the PDP is tiny. This 
also agrees with the conjecture that the PDP is due to perturbations of 
suitable fields (e.g. light scalar fields) arising from their quantum 
fluctuations during a period of inflation.

\section{The inflationary paradigm}
Before discussing the mechanisms through which the quantum fluctuations of
suitable fields can generate the curvature perturbation from inflation, it is
necessary to briefly present the so-called inflationary paradigm. This is the
typical manner in which inflation is modelled in particle cosmology. According
to the inflationary paradigm the Universe undergoes inflation when dominated by
the potential density of a scalar field, which is called the inflaton field.

We return to the Lagrangian of Eq.~(\ref{Lphim}) but this time instead of 
the mass term we consider a generic function $V(\phi)$, which corresponds to
the potential density of the scalar field. In this case, Eq.~(\ref{eomhom})
has the form
\begin{equation}
\ddot\phi+3H\dot\phi+V'(\phi)=0\,,
\label{eomhomV}
\end{equation}
where the prime denotes derivative with respect to the scalar field $\phi$.
The above equation is similar to the equation of motion of a body sliding down
the potential $V$ and subject to a friction term determined by the rate of the 
Universe expansion $H$. Based on this analogy we will consider the field as
rolling down the potential in field space.

Now, according to the inflationary paradigm, for inflation to take place
the Universe must be dominated by the potential density of the inflation field,
i.e. the kinetic density $\rho_{\rm kin}(\phi)$ of the field needs to be
subdominant:
\begin{equation}
V(\phi)\gg\rho_{\rm kin}(\phi)\equiv\frac12\dot\phi^2.
\end{equation}
This means that the field hardly moves and, therefore, its potential density
remains roughly constant, providing thereby the effective cosmological
constant \mbox{$\Lambda_{\rm eff}\equiv V/m_P^2$} needed for inflationary 
expansion. When the condition \mbox{$\rho_{\rm kin}\ll V$} is valid it is said
that the field undergoes slow-roll and its variation (motion in field space) is
overdamped by the friction term in Eq.~(\ref{eomhomV}). In this case one can 
ignore the acceleration term in Eq.~(\ref{eomhomV}) which is then recast as
\begin{equation}
3H\dot\phi\simeq -V'(\phi)\,.
\label{SR}
\end{equation}
Because the roll of the field is overdamped, the value of $\dot\phi$ is very 
small, which means that the slope of the potential needs to be very small as
well, according to the slow-roll equation above. Hence, the inflaton field 
corresponds to a flat direction in field space.

\begin{figure}
\begin{center}
\includegraphics[width=4in]{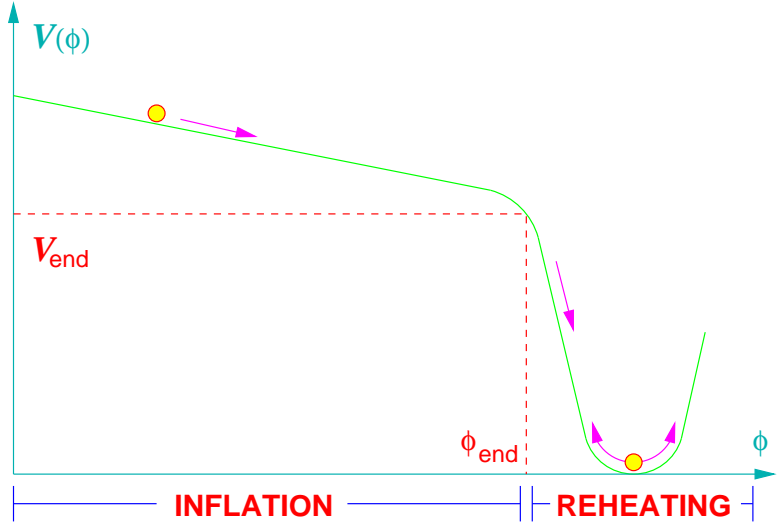}
\end{center}
\caption{\label{infpot}
Schematic representation of the inflaton potential $V(\phi)$. During inflation 
the inflaton field $\phi$ (represented by a circle) rolls down its flat 
potential in field space until it reaches a critical value $\phi_{\rm end}$ 
(corresponding to 
\mbox{$V_{\rm end}\equiv V(\phi_{\rm end})$}) when slow-roll is terminated and 
inflation ends. Afterwards the field undergoes rapid oscillations around its
VEV, corresponding to the minimum of the potential. The field decays while 
oscillating reheating thereby the Universe.}
\end{figure}

Inflation ends when the slow-roll condition is violated, i.e. at some critical 
value $\phi_{\rm end}$ of the inflaton field,
the potential becomes steep and curved such that \mbox{$\rho_{\rm kin}\sim V$}.
Afterwards, the inflaton condensate oscillates around its vacuum expectation 
value (VEV). Such coherent oscillations correspond to massive particles 
(inflatons)\footnote{with zero momentum since the field is homogeneous}, which
eventually decay into the standard model particles that comprise the thermal
bath of the Hot Big Bang (see Fig.~\ref{infpot}). 
This process is called Reheating and, if it occurs in
a perturbative manner, it is usually modelled by adding a phenomenological 
decay term in the field equation
\begin{equation}
\ddot\phi+3H\dot\phi+\Gamma\dot\phi+V'(\phi)=0\,,
\label{eomhomG}
\end{equation}
where $\Gamma$ stands for the decay rate of the inflaton field. Reheating 
occurs when the density of inflation is transferred to the newly formed thermal
bath, whose temperature is \mbox{$T_{\rm reh}\sim\sqrt{\Gamma m_P}$}, called
the reheating temperature.

\section{Mechanisms for the formation of the curvature perturbation}
We are ready now to discuss some ways that the perturbations of a light scalar 
field can give rise to the curvature perturbation in the Universe, which 
corresponds to the PDP that is responsible for structure formation.

\subsection{The inflaton hypothesis}
The traditional mechanism through which the curvature perturbation is generated
employs the inflaton field itself for the job. This, so-called inflaton
hypothesis, simply assumes that the field responsible for the formation of the
curvature perturbation is the same field which determines the dynamics of 
inflation, i.e. the inflaton field.

\begin{figure}
\begin{center}
\includegraphics[width=4in]{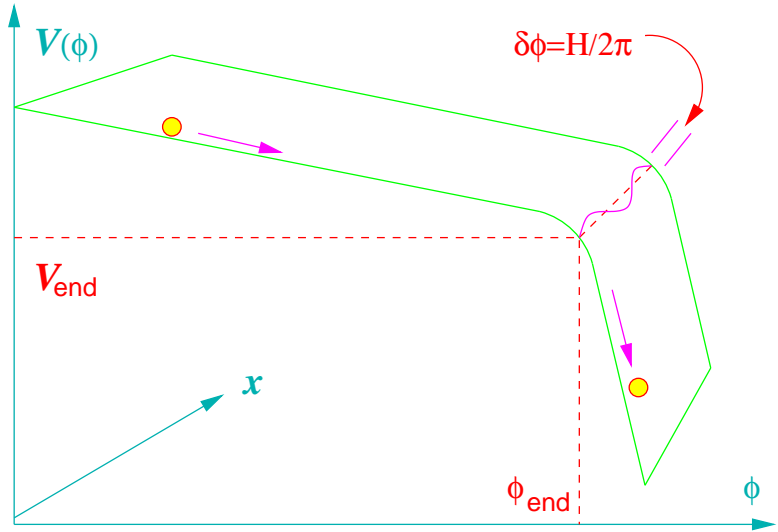}
\end{center}
\caption{\label{infend}
Schematic representation of the way that the curvature perturbation is 
generated in the inflaton hypothesis. The inflaton field perturbations vary 
its expectation value typically by \mbox{$\delta\phi=H/2\pi$}. This means
that, while the inflaton field (represented by a circle) rolls down its 
potential in field space, its value is different throughout coordinate space, 
represented by the $x$-labelled axis. Thus, in some locations the rolling
inflaton is ahead compared to others, which means that, in these locations, 
it will reach the critical value $\phi_{\rm end}$ that terminates inflation,
somewhat earlier. At a given time the inflaton value is represented by the 
wiggly line which crosses the \mbox{$\phi=\phi_{\rm end}$} line in several 
places. Therefore, the Universe inflates more in some places than in others, 
because, at the given time that corresponds to the wiggly line shown, in some 
locations inflation has ended while in others it still continues.}
\end{figure}

As mentioned in the previous section, to undergo slow-roll the inflaton must
correspond to a flat direction in field space and therefore is characterised by
\mbox{$V''\ll H^2$}, i.e. it is a light scalar field. This means that, through
the particle production process, it obtains a scale-invariant superhorizon 
spectrum of perturbations. As a result, the perturbed inflaton field will reach
the critical value $\phi_{\rm end}$ which terminates inflation at slightly 
different 
times at different points in space. Thus, inflation will continue a little bit 
more in some locations than in others (see Fig.~\ref{infend}). 
This is the cause of the generation of 
the curvature perturbation $\zeta$. Indeed, it can be shown that the latter is
determined by the difference in logarithmic expansion between the uniform
density and the spatially flat slices, i.e. \mbox{$\zeta=\delta(\ln a)$}
\cite{dN}.

The PDP in this case is given by
\begin{equation}
\left.\frac{\delta\rho}{\rho}\right|_H=
\left.\frac{H^2}{5\pi\dot\phi}\right|_*\;,
\label{drH}
\end{equation}
where the asterisk denotes the epoch of horizon exit of the inflaton 
perturbations. Since in the inflaton hypothesis a single degree of freedom 
(the inflaton) determines both the dynamics of inflation and the PDP, the 
latter
can be written in terms only of the characteristics of the inflaton potential.
Indeed, Eq.~(\ref{drH}) can be recast as \cite{book}
\begin{equation}
\left.\frac{\delta\rho}{\rho}\right|_H
=\frac{1}{5\sqrt 3\pi}\left.\frac{V^{3/2}}{m_P^3|V'|}\right|_*\;,
\end{equation}
which again is to be evaluated at horizon exit. Observational constraints on
the amplitude of the PDP (cf. Eq.~(\ref{drobs})) suggest that, if 
$\varepsilon$ is not extremely small, the energy scale of inflation 
is comparable to that of grand unification, i.e. 
\mbox{$V^{1/4}\sim 10^{15-16}\,$GeV}. This seems a natural scale to introduce 
new physics but it turns out that, for single field models, it requires
fine-tuning of model parameters.

The spectral index in the inflaton hypothesis is \cite{book}
\begin{equation}
n_s=1+2\eta_\phi-6\varepsilon\,,
\end{equation}
where \mbox{$\eta_\phi\equiv m_P^2V''/V$} corresponds to the curvature of the
potential along the direction of the inflaton field. If 
\mbox{$V(\phi)=\frac12 m_\phi^2\phi^2$} as in Eq.~(\ref{Lphim}), then 
\mbox{$\eta_\phi=\frac13(\frac{m_\phi}{H})^2$} as discussed after 
Eq.~(\ref{Pdphi})\footnote{we considered also the Friedman equation with 
\mbox{$\rho\simeq V$} since we have potential domination during inflation.}.

Under the inflaton hypothesis the generated non-Gaussianity is expected to be
negligible 
with \mbox{$\fnl\ll 1$}, i.e. below the limit of 
observability. This means that if non-Gaussianity is indeed observed in the PDP
then all the single field inflation models are going to be falsified.

The beauty and the curse of the inflaton hypothesis is that it relies on a 
single degree of freedom, namely the inflaton field, to account for all the
problems that inflation aims to address, i.e. provide the period of 
accelerated expansion that deals with the horizon and flatness problems and
also generate the appropriate spectrum of curvature perturbations, which agrees
with observational constrains on its amplitude and spectral index. As a result,
models of single field inflation are typically overconstrained and suffer from 
substantial fine-tuning. This is why alternative hypotheses have been put
forward for the generation of the PDP from inflation.

\subsection{The curvaton hypothesis}
This hypothesis assumes that the field responsible for the formation of the
curvature perturbation has nothing to do with the dynamics of inflation, i.e. 
it is {\em other} than the inflaton field. This scalar field is called curvaton
$\sigma$ \cite{curv} (see also 
Refs.~\cite{enqv}\cite{moroi}\cite{sylvia}\cite{linde}). 
In order for the curvaton to play this role it needs to obtain a
superhorizon spectrum of perturbations during inflation. Thus, it needs to be
a light field during inflation so that it can undergo particle production. 

It is evident that by introducing another degree of freedom, the fine-tuning
problems of inflation model-building are substantially improved
\cite{liber}\cite{liber+}. However, it
must be stressed that the curvaton is not necessarily a new ad hoc addition to 
the theory, which is introduced by hand. Indeed, the fact that the curvaton is 
not connected to the inflaton sector allows it to correspond to physics at 
energy scales much smaller than that of inflation (e.g. the TeV scale which is 
accessible by LHC). Consequently, there exist a number of candidates for the 
curvaton 
which correspond to realistic fields already present in simple extensions of 
the Standard Model. Prominent examples are a right-handed neutrino 
\cite{john}, an MSSM flat direction \cite{mssm1}\cite{mssm2}
or the so-called orthogonal axion in supersymmetric 
realisations of the Peccei-Quinn symmetry \cite{chun}\cite{ortho}. 

Under the curvaton hypothesis the curvature perturbation is given by 
\cite{unga}
\begin{equation}
\zeta=\hat\Omega_\sigma\zeta_\sigma\;,
\label{zcurv}
\end{equation}
where \mbox{$\hat\Omega_\sigma\equiv\frac{3\Omega_\sigma}{4-\Omega_\sigma}
\simeq\Omega_\sigma$}, where 
\mbox{$\Omega_\sigma\equiv(\rho_\sigma/\rho)_{\rm dec}$} is the density
parameter of the curvaton field at the time of its decay after 
inflation\footnote{$\hat\Omega_\sigma$ is also denoted as $r$ in much of the 
literature.}. In the above $\zeta_\sigma$ is the curvature perturbation 
attributed to the curvaton field, which is determined by the fractional 
perturbation of the field itself:\footnote{in spatially flat hypersurfaces}
\begin{equation}
\zeta_\sigma\equiv-H\frac{\delta\rho_\sigma}{\dot\rho_\sigma}=
\frac13\frac{\delta\rho_\sigma}{\rho_\sigma}=
\frac23\frac{\delta\sigma}{\sigma}\simeq
\frac23\left.\frac{\delta\sigma}{\sigma}\right|_*=\frac{H_*}{3\pi\sigma_*}\,,
\label{zs}
\end{equation}
where we used that \mbox{$\delta\sigma_*=H_*/2\pi$} and we
assumed that near its decay the curvaton density is 
\mbox{$\rho_\sigma\sim m_\sigma^2\sigma^2$}, where \mbox{$m_\sigma\ll H_*$}
is the mass of the curvaton field. The spectral index of the PDP in this case
is \cite{unga}
\begin{equation}
n_s=1+2\eta_\sigma-2\varepsilon\,,
\label{nscurv}
\end{equation}
where 
$\eta_\sigma$ corresponds to
the curvature of the potential along the curvaton direction.

\begin{figure}
\begin{center}
\includegraphics[width=4in]{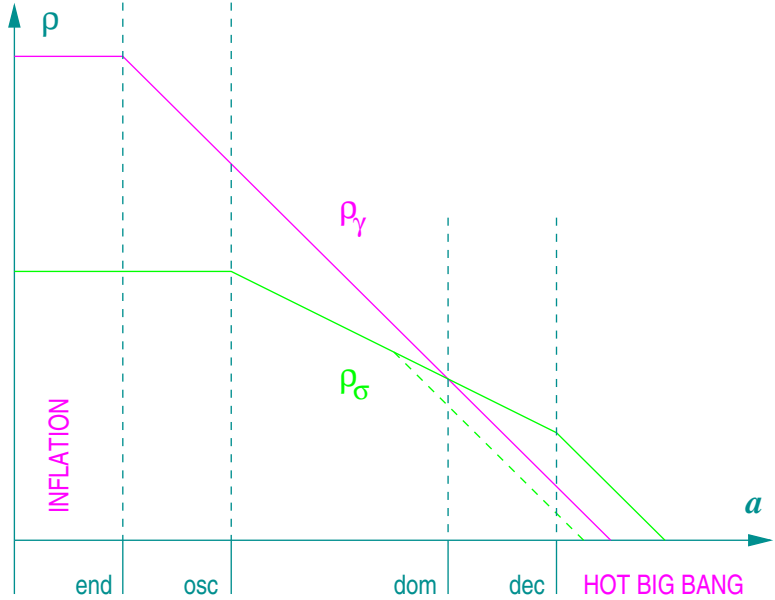}
\end{center}
\caption{\label{curvevol}
Log-log plot of the evolution of the inflaton energy density which decays
into radiation $\rho_\gamma$ (purple line) 
at the end of inflation (denoted by `end') 
and the curvaton energy density $\rho_\sigma$ (green line)
(prompt reheating is assumed). 
During inflation, the curvaton density is negligible.
After inflation \mbox{$\rho_\gamma\propto a^{-4}$}. In contrast,
\mbox{$\rho_\sigma$} remains constant (the curvaton is frozen at some value 
$\sigma_*$) until \mbox{$m\sim H(t)$}, when the curvaton unfreezes and begins
oscillating, after which time (denoted `osc') 
\mbox{$\rho_\sigma\propto a^{-3}$}. At some moment (denoted `dom') the curvaton
density dominates the Universe (\mbox{$\hat\Omega_\sigma=\Omega_\sigma=1$})
until, some time later (denoted `dec') when
it decays into the thermal bath of the Hot Big Bang. The dashed slanted line 
depicts the possibility that the curvaton decays before domination 
(\mbox{$\hat\Omega_\sigma=\frac34\Omega_\sigma\ll 1$}), when substantial 
non-Gaussianity can be generated.%
}
\end{figure}

Since by definition the curvaton should not affect the dynamics of inflation,
during inflation we expect \mbox{$\rho_\sigma\ll\rho$}. Thus the density
parameter of the curvaton is extremely small and its contribution to the 
overall
curvature perturbation (cf. Eq.~(\ref{zcurv})) is also small. For the curvaton
to significantly contribute to $\zeta$ we need to consider the period after 
inflation, when its contribution to the density can increase.

In the simplest case when \mbox{$V(\sigma)\simeq\frac12 m_\sigma^2\sigma^2$}
the equation of motion for the curvaton is of the same form as 
Eq.~(\ref{eomhom}). Since during inflation the field is light and 
\mbox{$m_\sigma\ll H$}, the curvaton is overdamped and remains frozen in 
some value $\sigma_*$. After inflation, however, the Hubble parameter decreases
in time \mbox{$H(t)\propto 1/t$}. As a result, there will be a moment when
\mbox{$m_\sigma\sim H(t)$}, at which time the curvaton condensate will unfreeze
and will begin coherent oscillations around its VEV. These oscillations 
correspond to massive particles (curvatons) whose density is diluted as
\mbox{$\rho_\sigma\propto a^{-3}$}, which is less drastic that the density
of the radiation background\footnote{This is the radiation that was generated 
by the decay of the inflaton field after the end of inflation.} 
\mbox{$\rho_\gamma\propto a^{-4}$}. Thus, the oscillating curvaton has a chance
to dominate (or nearly dominate) the Universe before its decay 
(see Fig.~\ref{curvevol}). Thus, at 
curvaton decay \mbox{$\hat\Omega_\sigma$} can be as large as unity and the 
curvaton contribution to $\zeta$ can be substantial. Consequently, the curvaton
imposes its curvature perturbation onto the Universe at (or near) its 
domination. 

Because of its spectrum of perturbations $\delta\sigma$ the amplitude of the
curvaton oscillations is also perturbed, i.e. it is larger in some places than 
in others. This means that the density of the oscillating field is perturbed 
too and the field will dominate the Universe at different times at 
different locations. This is what generates the curvature perturbation 
(cf. Eq.~(\ref{zs})).

In contrast to the inflaton hypothesis, under the curvaton hypothesis 
non-Gaussianity can be substantial. Indeed, in this case \cite{unga}
\begin{equation}
\fnl=\frac{5}{4\hat\Omega_\sigma}\,,
\label{fnlcurv}
\end{equation}
which can be large if the decay of the curvaton happens before domination
when \mbox{$\Omega_\sigma\ll 1$}. In fact, the WMAP observations in 
Eq.~(\ref{fnlobs}) set the lower bound  \mbox{$\Omega_\sigma\gsim 0.01$}.

\subsection{Other mechanisms}

There are numerous other ways to generate the PDP from a superhorizon spectrum
of scalar field perturbations. This section briefly reviews two of them, namely
the inhomogeneous reheating and the end of inflation mechanisms. As with the 
curvaton mechanism, these mechanisms assume that the contribution to the 
curvature perturbation from the inflaton field itself is negligible.

\subsubsection{Inhomogeneous Reheating}
This mechanism generates the PDP by assuming that the inflaton decay rate 
$\Gamma$ is modulated by another scalar field $\sigma$ 
\cite{modreh1}\cite{modreh2}. This scalar field is 
the one which undergoes particle production during inflation and which obtains
thereby a superhorizon spectrum of perturbations.

In this scenario we can ignore the perturbations of the inflaton field.
Thus, we can consider that inflation is terminated at the same time throughout 
space\footnote{at least in the observable Universe.}. The inflaton then begins
its coherent oscillations around its VEV until it decays when 
\mbox{$\Gamma\simeq H(t)$}. However, since in this case 
\mbox{$\Gamma=\Gamma(\sigma)$} and $\sigma$ is perturbed in space, the decay 
rate is different at different locations so that the Hot Big Bang begins at
different times. This is what generates the curvature perturbation which 
corresponds to the PDP. Roughly speaking we have
\begin{equation}
\frac{\delta\rho}{\rho}\sim\frac{\delta\Gamma}{\Gamma}\sim
\frac{\delta\sigma}{\sigma}\,.
\end{equation}
In this scenario the spectral index is again given by Eq.~(\ref{nscurv}). 
Non-Gaussianity in this model is quantified as \cite{zalda}
\begin{equation}
\fnl=5\left(\frac{\Gamma''\Gamma}{\Gamma'^2}-1\right),
\end{equation}
where now the prime denotes derivative with respect to $\sigma$. If $\Gamma$ 
has a power-law dependence on $\sigma$ then \mbox{$\fnl={\cal O}(1)$}, which is
marginally observable.

\subsubsection{End of inflation}\label{einf}
This mechanism applies to a particular type of inflation model, the so-called
hybrid inflation. Therefore, before discussing the mechanism a brief summary of
hybrid inflation is necessary.

Hybrid inflation couples the inflaton field $\phi$ to another scalar field 
$\psi$, called the waterfall field, in such a way that inflation is terminated 
by a phase transition which sends the waterfall to its VEV \cite{hybrid}. 
Since inflation is
likely to be at the scale of grand unification, in many cases the waterfall 
field is assumed to be the Higgs field of a Grand Unified Theory (GUT). Thus, 
inflation is 
terminated by the breaking of grand unification, i.e. the GUT phase transition.

The scalar potential for hybrid inflation has the following form
\begin{equation}
V(\phi,\psi)=\frac14\lambda(\psi^2-M^2)^2+\frac12g\phi^2\psi^2+V_\phi(\phi)\,
\label{Vhybrid}
\end{equation}
where $M$ is the GUT energy scale, $\lambda$ is the self-coupling of the 
waterfall field and $g$ is the interaction coupling between the waterfall and 
the inflaton. The potential $V_\phi(\phi)$ is responsible for the slow-roll of
the inflaton and its precise form is not relevant to this 
discussion\footnote{In supersymmetric versions of hybrid inflation the 
slow-roll potential is provided by radiative corrections and it is of the form
\mbox{$V_\phi\propto\ln\phi$} \cite{susyhybrid}\cite{laza}.}. 
From the above potential it is evident that 
the effective mass-squared of the waterfall field is
\begin{equation}
m_\psi^2=g\phi^2-\lambda M^2+\lambda\phi^2.
\label{mpsi}
\end{equation}
The above implies that there is a critical value of the inflaton
\begin{equation}
\phi_c=\sqrt{\frac{\lambda}{g}}M
\label{phic}
\end{equation}
such that if \mbox{$\phi\gg\phi_c$} then \mbox{$m_\psi^2>0$}
and \mbox{$\psi\rightarrow 0$}. In this case, Eq.~(\ref{Vhybrid}) becomes
\mbox{$V=\frac14\lambda M^4+V_\phi$}. The constant contribution to the scalar 
potential provides the effective cosmological constant for (quasi)de Sitter 
inflation: \mbox{$\Lambda_{\rm eff}\sim\lambda M^4/m_P^2$}. If however, 
\mbox{$\phi<\phi_c$} then \mbox{$m_\psi^2<0$} and a phase transition occurs
which results in \mbox{$\psi\rightarrow M$}, which gives the inflaton a large 
mass (through the interaction term) and so \mbox{$\phi\rightarrow 0$}. 
Assuming \mbox{$V_\phi(0)=0$}, after the phase transition 
\mbox{$V\rightarrow 0$} and inflation ends.

The end of inflation mechanism for the production of the PDP introduces an
extra coupling between the waterfall field and another scalar field $\sigma$
\cite{end}\cite{end+}. The scalar potential now becomes
\begin{equation}
V(\phi,\psi)=\frac14\lambda(\psi^2-M^2)^2+\frac12g\phi^2\psi^2+V_\phi(\phi)
+\frac12 h\sigma^2\psi^2\,
\label{Vhybrid+}
\end{equation}
where $h$ parametrises the strength of the interaction between $\sigma$ and the
waterfall field $\psi$. With this addition the effective mass-squared of the
waterfall field becomes
\begin{equation}
m_\psi^2=g\phi^2-\lambda M^2+\lambda\phi^2+h\sigma^2
\label{mpsi+}
\end{equation}
and the critical value which triggers the phase transition that ends inflation
is now
\begin{equation}
\phi_c(\sigma)=\left(\frac{\lambda}{g}M^2-\frac{h}{g}\sigma^2\right)^{1/2},
\label{phic+}
\end{equation}
i.e. it is modulated by the value of $\sigma$. Therefore, if $\sigma$ is light 
during inflation, then it undergoes particle production and obtains a 
superhorizon spectrum of perturbations with typical magnitude 
\mbox{$\delta\sigma=H/2\pi$}. This means that the value of $\phi_c$ is also 
perturbed. Hence, the phase transition which terminates inflation occurs 
earlier in some parts of the Universe than in others depending on when the 
inflaton reaches the critical value $\phi_c$. Consequently, the Universe 
inflates more in some locations than in others and this generates the curvature
perturbation.
As in the curvaton scenario, the spectral index in this case is
given by Eq.~(\ref{nscurv}) \cite{laila}.

\section{Cosmic vector fields and the curvature perturbation}

Tantalising evidence exists of a preferred direction in the CMB temperature
perturbations. In particular, the low multipoles in the CMB appear to be 
aligned \cite{AoE}\cite{AoE+} at a level which is statistically rather 
improbable \cite{eriksen}\cite{hansen}. A preferred 
direction in the CMB cannot be accounted for if inflation and the generation
of the PDP is due to scalar fields only. Moreover, despite their abundance in
theories beyond the Standard Model, scalar fields have not been observed as 
yet. 
If the Higgs field is not found in the LHC, the credibility of the ubiquitous 
usage of scalar fields in cosmology will be shaken.

Until recently only scalar fields have been considered both as responsible for
the dynamics of inflation and also for the generation of the observed PDP.
However, in the pioneering work of Ref.~\cite{vecurv} the possibility that a 
vector 
field contributes in the generation of the PDP was first considered. Since 
then, a number of attempts have been made to investigate the role and the 
implications of cosmic vector fields in the generation of the PDP
(see Ref.~\cite{varkin} and references therein). 

Why vector fields were not considered originally for the generation of 
the PDP? After all, these are fields which are similar to the massive gauge 
bosons, which have indeed been observed in LEP. The main difficulty had to do
with the inherent anisotropic nature of vector fields. Inflation would 
homogenise a vector field condensate, and a homogeneous vector field picks up 
a preferred direction in space. Thus, if this vector field condensate were to 
dominate the density of the Universe (so that it can affect the expansion and 
generate the PDP) it was felt that the anisotropic stress generated would lead 
to strongly anisotropic expansion which would be impossible to reconcile with
the predominant isotropy of the CMB. However, as is discussed below, this
difficulty can be circumvented in the vector field remains subdominant during
inflation. Still, if (in analogy to scalar fields) light vector fields were 
needed for a scale invariant spectrum then a more subtle problem arises. 
Massless Abelian vector fields are conformally invariant and they cannot 
undergo 
particle production during inflation\footnote{They view the Universe expansion 
as a conformal transformation to which they are insensitive.}. Thus, for light
vector fields particle production is expected to be suppressed. Nevertheless,
there are numerous mechanisms which break the conformality of vector fields so
that particle production can be efficient. These mechanisms are model dependent
which means that they can have distinct observational signatures as is 
discussed below.

\section{Particle production of vector fields during inflation}
Suppose that there is a suitable theory which breaks the vector field 
conformality during inflation. How are we to study particle production of the
vector field? We follow the same recipe as in Sec.~\ref{pp} assuming that the 
inflationary expansion remains isotropic.

Firstly we perturb the vector field around its homogenised value
\mbox{$A_\mu=A_\mu(t)+\delta A_\mu(\vec x,t)$}. Then we Fourier transform the
perturbations 
\mbox{$\delta A_i(\vec k)=\int\delta A_i(\vec x)e^{-i\vec k\cdot\vec x}d\vec x
$} and we obtain the equations of motion for the Fourier components in the
theory which we are considering\footnote{We focus on the spatial components
of the cosmic vector field, as they are the ones which would generate 
anisotropy.
Also, the temporal component of a homogeneous massive Abelian vector field is 
zero.}. Then we promote the perturbations of the vector field into quantum
operators
\begin{equation}
\delta\hat A_i(\vec x,t)=\int\frac{d^3k}{(2\pi)^3}\sum_\lambda\left[
\hat e_i^\lambda\hat a_\lambda(\vec k)\delta{\cal A}_k(k,t)
e^{i\vec k\cdot\vec x}+
\hat e_i^{\lambda\,*}\hat a_\lambda^\dagger(\vec k)\delta{\cal A}^*_k(k,t)
e^{-i\vec k\cdot\vec x}\right],
\label{vecexpan}
\end{equation}
where $\delta{\cal A}_k$ are the mode functions and we consider again canonical
quantisation 
\mbox{$[\hat a_\lambda(\vec k),\hat a_{\lambda'}^\dagger(\vec k')]=
(2\pi)^3\delta(\hat k-\hat k')\delta_{\lambda\lambda'}$}. In the above we have
introduced the polarisation vectors
\begin{equation}
\hat e^L\equiv\frac{1}{\sqrt 2}(1,i,0)\qquad
\hat e^R\equiv\frac{1}{\sqrt 2}(1,-i,0)\qquad
\hat e^\|\equiv(0,0,1),
\end{equation}
where $L,R$ denote the left and right transverse polarisations and $\|$
denotes the longitudinal 
polarisation\footnote{The index $\lambda$ in Eq,~(\ref{vecexpan}) runs over 
these values.}.

As with the scalar field case, the mode functions of the above expansion are
expected to satisfy the equations of motion for the Fourier components of the
vector field perturbations. The form of these equations depends on the theory
which breaks the conformality of the vector field. To solve them we again
consider vacuum boundary conditions in the subhorizon limit 
(\mbox{$k/aH\rightarrow\infty$}). They are \cite{stanis}:
\begin{equation}
\delta{\cal A}_k^{L,R}
\mbox{\raisebox{-1ex}{~$\stackrel{\mbox{\Large $\longrightarrow$}}{{
\mbox{\tiny $k\!/\!a\!H\!\!\rightarrow\!\!+\!\infty$}}}$~}}
\frac{e^{ik/aH}}{\sqrt{2k}}\quad{\rm and}\quad
\delta{\cal A}_k^\|
\mbox{\raisebox{-1ex}{~$\stackrel{\mbox{\Large $\longrightarrow$}}{{
\mbox{\tiny $k\!/\!a\!H\!\!\rightarrow\!\!+\!\infty$}}}$~}}
\gamma\,\frac{e^{ik/aH}}{\sqrt{2k}}\,.
\end{equation}
We see that the boundary condition is again the Bunch-Davies vacuum but the 
longitudinal component it is multiplied by the Lorentz boost factor
\mbox{$\gamma\equiv\frac{E}{m}=\sqrt{(\frac{k}{m})^2+1}$}, which takes us from
the frame with \mbox{$\vec k=0$} where all components are equivalent, to the
frame of momentum $\vec k$.\footnote{Note that for a massless vector field
the longitudinal component is unphysical.}

After solving the equations of motion we evaluate the solutions in the 
superhorizon limit (\mbox{$k/aH\rightarrow 0$}). Then we can obtain the power
spectrum for each polarisation as 
\mbox{$\calp_\lambda(k)\equiv\frac{k^3}{2\pi^2}|\delta{\cal A}_k^\lambda|^2$},
where $\delta{\cal A}_k^\lambda$ is evaluated in the superhorizon limit. 
Because we have three degrees of freedom there are three possibilities:


\medskip

Case A:
$\calp_\|\neq\calp_L\neq\calp_R\neq\calp_\|$

Case B:
$\calp_L=\calp_R\neq\calp_\|$

Case C:
$\calp_L=\calp_R=\calp_\|$

\medskip

Even though Case~A appears to be the most generic, in practice it is Case~B 
that is the most common because usually the theories which break the vector 
field
conformality are parity conserving. We have isotropic particle production only
in the special Case~C. 

\begin{figure}
\begin{center}
\begin{minipage}{100mm}
\mbox{\hspace{-2.8cm}%
{
\resizebox*{5cm}{!}{\includegraphics{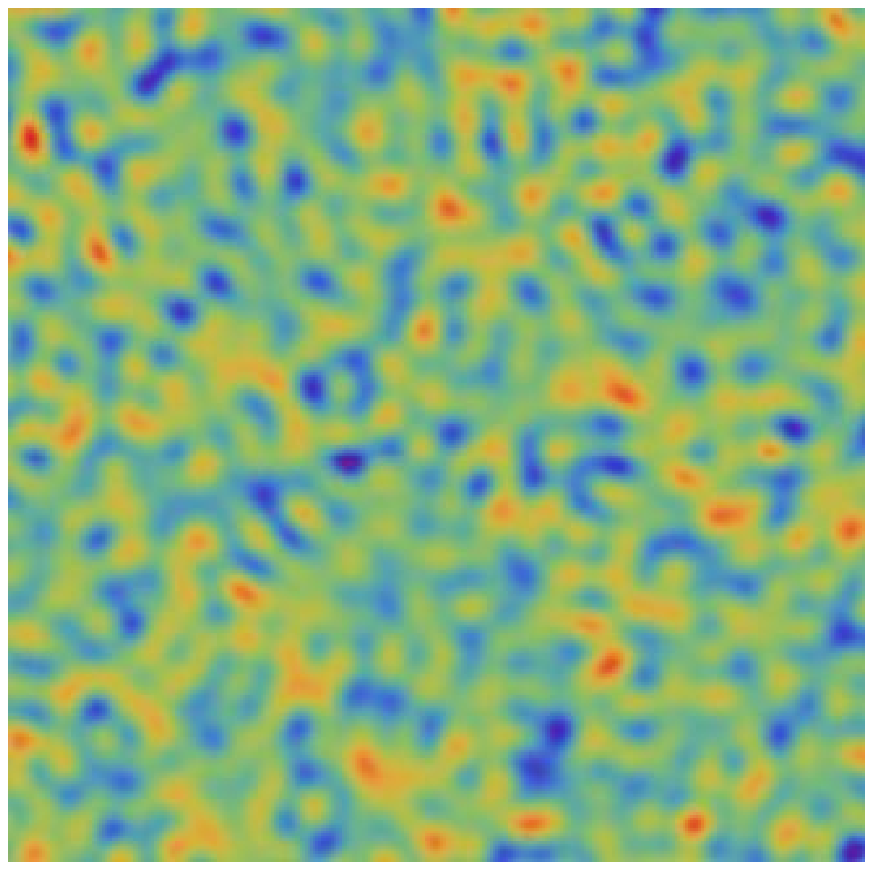}}}
{
\resizebox*{5cm}{!}{\includegraphics{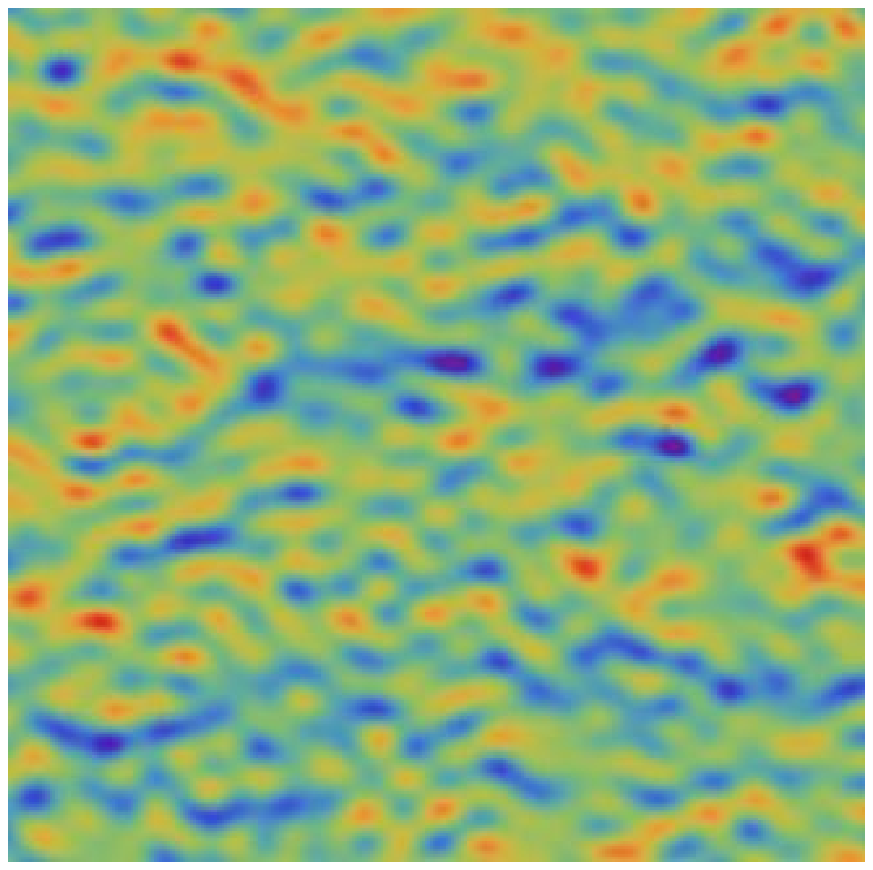}}}
{
\resizebox*{5cm}{!}{\includegraphics{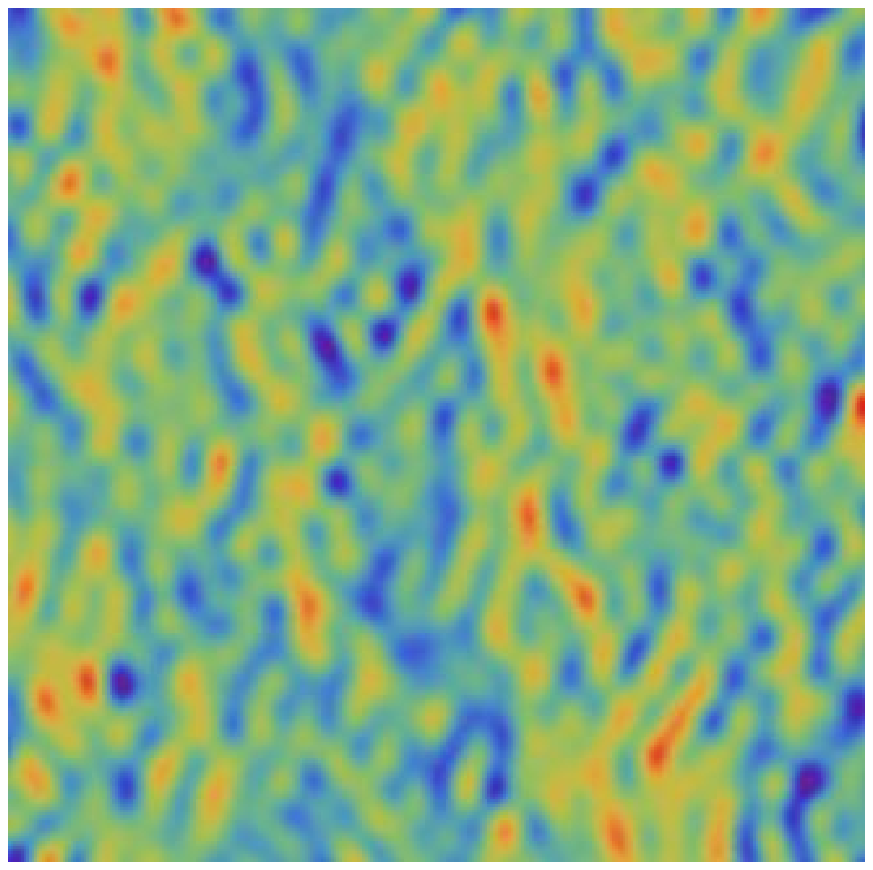}}}\\ %
}
\caption{%
Patterns in the CMB temperature perturbations which may arise from statistical
anisotropy in the CMB spectrum. The left panel shows
an isotropic signal, while the middle and right panels show patterns due to
statistical anisotropy along the vertical and horizontal direction 
respectively. The figure is taken from Ref.~\cite{MK}.}%
\label{CMBanis}
\end{minipage}
\end{center}
\end{figure}

What would happen if the spectrum of the produced perturbations of the vector
field affected the curvature perturbation? Since particle production is in 
general anisotropic we expect an anisotropic contribution to $\zeta$. This 
contribution can be parametrised as follows. For the power spectrum we can 
write \cite{acker}
\begin{equation}
\calp_\zeta(\vec k)=\calp_\zeta^{\rm iso}(k)
\left[1+g(\hat A\cdot\hat k)^2\cdots\right],
\label{stanisP}
\end{equation}
where \mbox{$\hat A\equiv \vec A/A$} and \mbox{$\hat k\equiv \vec k/k$} with
\mbox{$A\equiv|\vec A|$} and the ellipsis denotes higher order 
contributions. The anisotropic part of the spectrum corresponds to a new
observable, namely {\em Statistical Anisotropy}, which amounts to direction 
dependent patterns in the CMB \cite{stanis} (see Fig.~\ref{CMBanis}). 
Statistical anisotropy is quantified by the so-called
anisotropy parameter $g$ in Eq.~(\ref{stanisP}). The WMAP observations set a 
surprisingly weak upper bound on the anisotropy parameter: 
\mbox{$g\lssim 0.3$} \cite{firstg}\footnote{In Ref.~\cite{gbound}
statistical anisotropy with \mbox{$g=0.29\pm0.03$}
is claimed to be detected at a level of 9-$\sigma$ but the authors acknowledge
that, because the direction is suspiciously near the ecliptic plane, this is 
probably due to an unknown systematic error. Hence we treat this number as an 
upper bound.}. Thus we see that, at present, observations allow up to 30\% 
statistical anisotropy in the spectrum of the PDP. The forthcoming data of the 
Planck satellite will improve precision by an order of magnitude and are likely
to detect statistical anisotropy. If this will be so then a cosmic vector field
will have to be involved in the generation of the PDP during inflation.

In a similar manner one can parametrise the statistically anisotropic 
contribution from a vector field in the bispectrum. Indeed, we can write
\cite{fnlanis}\cite{cesar}
\begin{equation}
\fnl=\fnl^{\rm iso}\left(1+{\cal G}\hat A_\perp^2\cdots\right),
\label{fnlanis}
\end{equation}
where $\fnl^{\rm iso}$ denotes the isotropic part and $\hat A_\perp$ is the 
projection of $\hat A$ onto the plane of the three $\vec k$-vectors which are 
used to define the bispectrum. Note that, as 
non-Gaussianity has not been observed yet, there are no bounds on $\cal G$.
If \mbox{${\cal G}>1$} this means that non-Gaussianity is predominantly 
anisotropic (even though $\calp_\zeta$ is not). If non-Gaussianity is indeed 
observed (e.g. by the Planck satellite) and no angular modulation of $\fnl$
is found on the microwave sky then all models which predict 
\mbox{${\cal G}\gsim 1$} will be falsified. It is important also to stress that
the directions of statistical anisotropy in the spectrum and the bispectrum are
correlated (determined by the direction of $\hat A$), which is a smoking gun
for the contribution of a vector field to the PDP \cite{fnlanis}.

In the Cases~A and B the spectra of the superhorizon perturbations of the 
vector field are predominantly anisotropic. Since $\calp_\zeta$ is isotropic
at least at the level of 70\% the contribution of the vector field to the PDP
in these two cases
has to be subdominant and its significance is only the possibly observable 
statistical anisotropy in the spectrum and bispectrum. Thus, in the Cases~A 
and B we still require some other, isotropic source (such as a scalar field)
to provide the dominant contribution to the PDP. On the other hand, in Case~C
the perturbation spectra of the vector field are isotropic. In this case, the
vector field alone can generate $\zeta$ and no input from any scalar field is
necessary. Note that, in this case, the vector field may also produce 
statistical anisotropy if the $\calp_\lambda$ are not exactly the same but 
differ slightly, albeit by no more that 30\%.

\section{Models for particle production of cosmic vector fields}
In this section we discuss a couple of proposals for the generation of a
flat superhorizon spectrum of vector field perturbations.

\subsection{Non-minimal coupling to gravity}
Consider the following theory
\begin{equation}
{\cal L}=-\frac14F_{\mu\nu}F^{\mu\nu}+\frac12(m^2+\alpha R)A_\mu A^\mu,
\end{equation}
where \mbox{$F_{\mu\nu}=\partial_\mu A_\nu-\partial_\nu A_\mu$} is the
field strength of the Abelian vector field $A_\mu$, $m$ is its bare mass,
$\alpha$ is constant and $R$ is the Ricci scalar. From the above we see that 
the vector field has effective mass-squared \mbox{$m_A^2=m^2+\alpha R$}. 

Following the procedure outlined in the previous section, we obtain the 
following solution for the mode functions of the transverse perturbations of 
the physical vector field \cite{nonmin}%
\footnote{In a flat homogeneous and isotropic Universe
the spatial components of the physical (in contrast to comoving) vector field
are $A_i/a$.}
\begin{equation}
\delta{\cal A}^{L,R}_k=a^{-3/2}\sqrt{\frac{\pi}{H}}
\frac{e^{i\frac{\pi}{2}(\nu-\frac12)}}{1-e^{i2\pi\nu}}
[J_\nu(k/aH)-e^{i\pi\nu}J_{-\nu}(k/aH)].
\label{vectranssolu}
\end{equation}
The above is identical to Eq.~(\ref{phisolu}) with the crucial difference that
\mbox{$\nu\equiv\sqrt{\frac14-(\frac{m_A}{H})^2}$}. Using this solution we 
obtain the following expression for the power spectrum of the transverse
perturbations in the superhorizon limit
\begin{equation}
\calp_{L,R}=\frac{8\pi|\Gamma(1-\nu)|^{-2}}{1-\cos(2\pi\nu)}
\left(\frac{H}{2\pi}\right)^2\left(\frac{k}{2aH}\right)^{3-2\nu},
\end{equation}
which again is identical to Eq.~(\ref{Pphi}) but for the different value of 
$\nu$. From the above it is evident that a scale invariant spectrum is attained
for \mbox{$\nu\approx\frac32$}. This translates into a requirement for both
$\alpha$ and $m$ as follows. 

In a spatially flat, homogeneous and isotropic spacetime, the scalar curvature
is \mbox{$R=3(3w-1)H^2$}, where $w$ is the barotropic parameter of the content 
of spacetime. During inflation, \mbox{$w\simeq -1$} so that 
\mbox{$R\simeq -12\,H^2$}. This means that we can attain a scale invariant 
transverse spectrum if 1) \mbox{$\alpha=\frac16$} and 2) \mbox{$m\ll H$}, i.e.
the vector field is a light field with a negative effective 
mass-squared \cite{nonmin}\footnote{The non-minimal coupling to gravity 
\mbox{$\alpha=\frac16$} has a special property. It turns the conformally 
invariant massless Abelian vector field equivalent (in the sense that its field
equations are the same) to a (set of two) minimally coupled massless scalar 
field(s). If applied to a massless scalar field it renders it conformally 
invariant. In 
this respect it appears to reflect a deeper symmetry and therefore it may be 
a natural value for $\alpha$.}. Under these conditions the transverse spectra
are indeed scale-invariant and are given by 
\mbox{$\calp_{R,L}=(\frac{H}{2\pi})^2$}, exactly as a massless scalar field.

Employing the same conditions for $\alpha$ and $m$ one obtains the following
solution for the longitudinal component of the physical vector field 
\cite{stanis}:
\begin{equation}
\delta{\cal A}^\|_k=\frac{1}{\sqrt 2}
\left[\left(\frac{k}{aH}\right)-2\left(\frac{aH}{k}\right)+2i\right]
\frac{e^{ik/aH}}{\sqrt{2k}}\,.
\end{equation}
Clearly the above is totally different from Eq.~(\ref{vectranssolu}). In the 
superhorizon limit the power spectrum of the longitudinal component is
\mbox{$\calp_\|=2(\frac{H}{2\pi})^2=2\calp_{R,L}$}. This means that we are in 
Case~B since \mbox{$\calp_\|>\calp_{L,R}$} and
\mbox{$\calp_L=\calp_R$} because the theory is parity invariant.
Thus, particle production in this theory is anisotropic at a level of 100\%. 
Therefore, the vector field cannot alone be responsible for the PDP. Its 
contribution to $\zeta$ has to be subdominant and it can only be the source of
statistical anisotropy. Before concluding, we should point out that this model
has been criticised for suffering from instabilities (ghosts) 
\cite{peloso1}\cite{peloso2}
(see however Ref.~\cite{RA2}). 

\subsection{Varying kinetic function and mass}
Now, consider the theory 
\begin{equation}
{\cal L}=-\frac14fF_{\mu\nu}F^{\mu\nu}+\frac12m^2A_\mu A^\mu,
\end{equation}
where \mbox{$F_{\mu\nu}=\partial_\mu A_\nu-\partial_\nu A_\mu$},
\mbox{$f=f(t)$} is the kinetic function and we also consider
\mbox{$m=m(t)>0$} during inflation. The Maxwell-type kinetic term in 
combination with the positive mass-squared guarantees the stability of the 
model \cite{max}
and therefore is motivated even if the vector field is not a gauge boson.
Note also that a massive Abelian vector field is renormalisable even is it is
not a gauge field \cite{tikto}.

The solutions for the mode functions of the field perturbations are too
complicated to reproduce here. It suffices to say that scale invariance in the
transverse components requires the kinetic function to scale with the expansion
as \mbox{$f\propto a^{-1\pm 3}$} and also the physical vector field to be light
at horizon exit \mbox{$M\ll H$}, where \mbox{$M\equiv m/\sqrt f$}. Scale 
invariance for the longitudinal component additionally requires 
\mbox{$m\propto a$} \cite{varkin}\cite{noinsta}.

If the vector field is a gauge boson then $f$ is the gauge kinetic function,
which is related to the gauge coupling as \mbox{$f\sim 1/e^2$}. This means that
only the case when \mbox{$f\propto a^{-4}$} is acceptable because only then
the gauge field remains weakly coupled during inflation. Note that the
model is naturally realisable in supergravity theories where $f$ is a 
holomorphic function of the scalar fields of the theory \mbox{$f(\phi_i)$}
\cite{sugravec}.
In general, K\"{a}hler corrections to the scalar potential result in masses
of order $H$ for the scalar fields
\cite{randall}\footnote{This is the source of the famous 
$\eta$-problem of inflation, since slow-roll requires 
\mbox{$|\eta|=\frac13(\frac{m}{H})^2\ll 1$} for the inflaton.}, 
which are therefore expected to be fast-rolling
down the potential slopes during inflation, causing significant variation to 
$f$. Indeed, for a power-law dependence of the gauge kinetic function to the
scalar fields, it is easy to show that \mbox{$\dot f/f\sim H$} \cite{varkin}, 
i.e. $f$ has
a power-law dependence on $a$ as assumed in this model. Now, if $f$ is
modulated by the inflaton field itself, then it can be shown that, under 
fairly general conditions, the backreaction to the inflaton's roll renders the
scaling \mbox{$f\propto a^{-4}$} an attractor solution \cite{new}. 
Similarly, for
a gauge boson, $m$ can be modulated by a fast-rolling Higgs field with 
tachyonic mass \mbox{$m_H=2H$} \cite{varkin}.

Given the conditions \mbox{$f\propto a^{-1\pm 3}$} and \mbox{$m\propto a$}
the power spectra for the transverse and longitudinal components depend on
whether the vector field remains light until the end of inflation or not
(note that \mbox{$M=M(t)$}). In particular we find \cite{varkin}\cite{noinsta}
\begin{eqnarray}
M\ll H: & & \calp_L=\calp_R=
\left(\frac{H}{2\pi}\right)^2\quad{\rm and}\quad
\calp_\|=\left(\frac{H}{2\pi}\right)^2\left(\frac{3H}{M}\right)^2\\
 & & \nonumber\\
M\gg H: & & \calp_L=\calp_R=\calp_\|=
\frac12\left(\frac{H}{2\pi}\right)^2\left(\frac{3H}{M}\right)^2.
\end{eqnarray}
From the above we see that, if the vector field remains light until the 
end of inflation, \mbox{$\calp_\|\gg\calp_L=\calp_R$}, i.e. we are in Case~B.
Therefore, particle production is strongly anisotropic and the vector field
contribution to the PDP has to be subdominant but it can still give rise to
substantial statistical anisotropy. In contrast, if the field becomes heavy
by the end of inflation then particle production is isotropic and the vector 
field can be solely responsible for the generation of the PDP. Note that, 
because we need the field to be light when the cosmological scales exit the 
horizon, for it to become heavy we need \mbox{$\dot M>0$} during inflation, 
which is possible only in the case when \mbox{$f\propto a^{-4}$}. If 
\mbox{$f\propto a^2$} then \mbox{$M=\,$constant} and we have \mbox{$M\ll H$}
throughout inflation. But if \mbox{$f\propto a^{-4}$} then we have 
\mbox{$M\propto a^3$} so that we may end up having \mbox{$M\gg H$} at the end 
of inflation even though we started of with a light field at horizon exit.
In the latter case the vector field begins coherent oscillations a few 
exponential expansions (e-folds) before the end of inflation. Assuming that 
at the end of inflation \mbox{$f\rightarrow 1$} and the field becomes 
canonically normalised, we find that there is ample parameter space for Case~C
to be realised, namely: \mbox{$1<m/H<10^6$} \cite{varkin}\cite{noinsta}.

\section{The vector curvaton paradigm}
Through the examples in the previous section it is evident that a scale 
invariant spectrum of perturbations (isotropic or not) of a vector field can 
indeed be generated if one assumes some theory which appropriately breaks the 
conformality of the vector field. However, in order for these perturbations
to affect or even generate the PDP the vector field needs to affect the 
Universe expansion, i.e. its density should become dominant (or nearly 
dominant) at some point. But, even if particle production is isotropic the 
homogeneous zero-mode condensate is not. How can we avoid excessive anisotropic
stress when the vector field dominates the expansion? It turns out that this
is possible if the vector field plays the role of the curvaton.

Consider a minimally coupled massive Abelian vector field for which
\begin{equation}
{\cal L}=-\frac14F_{\mu\nu}F^{\mu\nu}+\frac12m^2A_\mu A^\mu,
\end{equation}
where \mbox{$F_{\mu\nu}=\partial_\mu A_\nu-\partial_\nu A_\mu$} and 
\mbox{$m=\,$constant$\,>0$}. 
It is clear that both models in the previous section 
eventually approach the above theory, when \mbox{$m^2\gg R\sim H^2$} and
also after the end of inflation when \mbox{$f\rightarrow 1$} and 
\mbox{$m=\,$constant}. The vector field condensate is homogenised by inflation
so that \mbox{$A_\mu=A_\mu(t)$}. In this case it can be shown that the temporal
component is zero and the spatial components satisfy the following equation
\cite{vecurv}
\begin{equation}
\ddot A_i+H\dot A_i+m^2A_i=0\,,
\label{veceom}
\end{equation}
which is similar to Eq.~(\ref{eomhom}) for the scalar field. The 
energy-momentum tensor of this theory can be written in the form \cite{vecurv}
\begin{equation}
T_\mu^\nu={\rm diag}(\rho_A, -p_\perp, -p_\perp, +p_\perp)\,,
\label{Tmn}
\end{equation}
where 
\begin{eqnarray}
\rho_A=\rho_{\rm kin}+V & & \rho_{\rm kin}\equiv-\frac14F_{\mu\nu}F^{\mu\nu}
\nonumber\\
 & \quad{\rm and}\quad & \\
p_\perp=\rho_{\rm kin}-V & & V\equiv-\frac12m^2A_\mu A^\mu.
\nonumber
\end{eqnarray}
Eq.~(\ref{Tmn}) is reminiscent of a perfect fluid with the crucial difference 
that the pressure in the longitudinal direction is of opposite sign compared to
the transverse pressure. Thus, it seems that if the homogeneous vector field 
were to dominate the Universe it would indeed generate excessive anisotropic 
stress. Therefore, we assume that the vector field remains subdominant during 
inflation so that (quasi)de Sitter expansion is not spoilt.

\begin{figure}
\begin{center}
\includegraphics[width=4in]{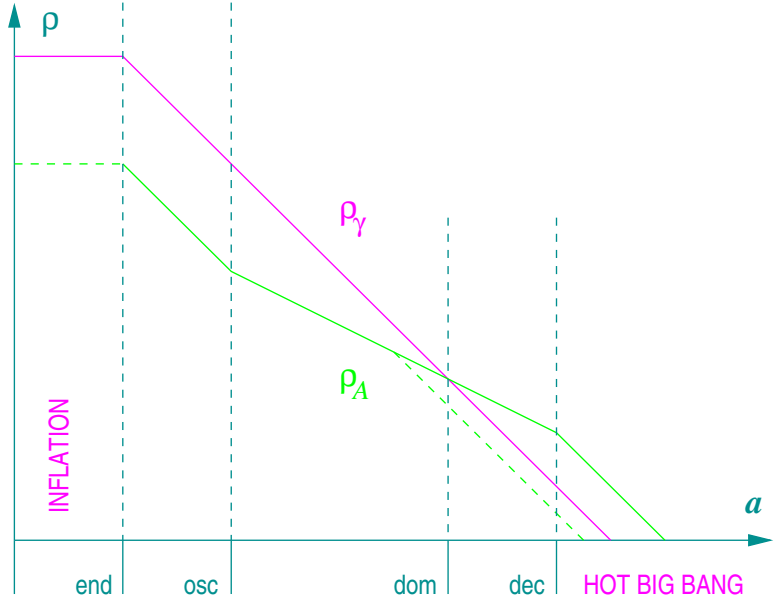}
\end{center}
\caption{\label{vecurvevol}
Log-log plot of the evolution of the inflaton energy density which decays
into radiation $\rho_\gamma$ (purple line) at the end of inflation 
(denoted by `end') and the vector curvaton energy density $\rho_A$ (green line)
(prompt reheating is assumed). 
During inflation, the vector curvaton density is negligible
(its evolution depends on the model which breaks its conformality).
After inflation \mbox{$\rho_\gamma\propto a^{-4}$}. Similarly,
\mbox{$\rho_A$} decreases also as radiation after inflation when the vector 
field is light (in contrast to the scalar curvaton case where $\rho_\sigma$ 
remains constant when the curvaton is light, cf. Fig.~\ref{curvevol}). 
However, when \mbox{$m\sim H(t)$}, the vector curvaton becomes heavy and begins
oscillating, after which time (denoted `osc') 
\mbox{$\rho_A\propto a^{-3}$}. At some moment (denoted `dom') the 
vector curvaton density dominates the Universe 
until, some time later (denoted `dec') when
it decays into the thermal bath of the Hot Big Bang. As the vector curvaton
is rapidly oscillating it does not give rise to anisotropic 
stress at domination. The dashed slanted line 
depicts the possibility that the vector curvaton decays before domination 
(\mbox{$\hat\Omega_A\ll 1$}), when substantial 
non-Gaussianity can be generated.%
}
\end{figure}

After inflation the Hubble parameter decreases until \mbox{$m>H(t)$}. When this
happens one can ignore the friction term in Eq.~(\ref{veceom}), which therefore
suggests that the vector field condensate starts rapid (quasi)harmonic 
oscillations. It is easy to show that, during a Hubble time, on average
\mbox{$\overline{\rho_{\rm kin}}\approx\overline V$}, which means that,
over many oscillations, the average pressure is zero:
\mbox{$\overline{p_\perp}=0$} \cite{vecurv}. 
Thus the oscillating vector field condensate 
behaves as pressureless, {\em isotropic} matter. Its density, therefore, scales
as \mbox{$\rho_A\propto a^{-3}$} which is not as drastic as the radiation
background, which scales as \mbox{$\rho_\gamma\propto a^{-4}$}. Therefore, 
the oscillating vector field can gradually increase its density parameter and
come to dominate (or nearly dominate) before its decay 
(see Fig.~\ref{vecurvevol}). Because it is 
isotropic, it dominates without causing any excessive anisotropic stress, so
that the Universe expansion remains isotropic \cite{vecurv}.

From the above we see that a massive Abelian vector field can follow the 
curvaton scenario and play the role of the curvaton without problem. Indeed, 
the perturbations of the vector field imply a perturbation in its local 
density $\rho_A(\vec x)$, which means that in some locations it dominates the 
Universe earlier than in others. This is how it can generate the curvature 
perturbation according to the curvaton mechanism. It is important to note 
that, since $\rho_A$ is a scalar quantity, the curvature perturbation generated
is scalar and not vector in nature. Also, note that the perturbations of the
vector field $\delta A_i$ follow a similar equation to Eq.~(\ref{veceom}), 
which means that they too are rapidly oscillating and do not introduce 
anisotropic stress. Thus, at domination
the perturbations do not cause anisotropic expansion 
either, not even in a small scale. If particle production of the vector field
is anisotropic then there are direction dependent patterns in the amplitude of 
the oscillating zero mode that lead to statistical anisotropy in the PDP
\cite{stanis}.

In Ref.~\cite{fnlanis} statistical anisotropy in the bispectrum in the vector 
curvaton model was investigated. It was found that it manifests itself only 
quadratically, i.e. the expansion in Eq.~(\ref{fnlanis}) is truncated to
\mbox{$\fnl=\fnl^{\rm iso}\left(1+{\cal G}\hat A_\perp^2\right)$}. $\cal G$ was
evaluated in the two models discussed in the previous section. In the 
non-minimally coupled to gravity model \mbox{${\cal G}^{\rm eql}=\frac98$} in 
the equilateral configuration, whereas \mbox{${\cal G}^{\rm sqz}=1$} in the 
squeezed configuration. This means that the angular modulation of $\fnl$
is prominent and should be detected if non-Gaussianity is found.
In the varying kinetic function and mass model
\mbox{${\cal G}^{\rm eql}=\frac18(\frac{3H}{M})^4\gg
{\cal G}^{\rm sqz}=(\frac{3H}{M})^2\gg 1$} when \mbox{$M\ll H$}, i.e. if the 
field remains light until the end of inflation. Here we see that 
non-Gaussianity is predominantly anisotropic. Should no angular modulation of 
$\fnl$ be observed this possibility will be excluded. However, if 
\mbox{$M\gg H$}, i.e. if the field becomes massive by the end of inflation, 
particle production is isotropic (Case~C) and 
\mbox{$\fnl=\fnl^{\rm iso}=5/4\hat\Omega_A$}, which is identical to the scalar 
curvaton case (cf. Eq.~(\ref{fnlcurv})).\footnote{$\hat\Omega_A$ is defined in 
the same way as $\hat\Omega_\sigma$.}

The vector curvaton is an elegant mechanism for vector fields to contribute of
even generate the PDP since it does not need to couple the fields to the 
inflaton sector. However, it is by no means the only way that a vector field 
can affect the curvature perturbation. For example, in Ref.~\cite{yoko}, the 
end of inflation mechanism is employed (see Sec.~\ref{einf}), where, instead of
coupling the waterfall field $\psi$ of hybrid inflation to a scalar field 
$\sigma$, the authors considered introducing a coupling of the form
\mbox{$\Delta V=\frac12hA_\mu A^\mu\psi^2$}. This model too produces distinct 
observational signatures. For example, in this model
\mbox{$\fnl=\fnl^{\rm iso}(1+{\cal G}\hat A_\perp^2+
{\cal G}'\hat A_\perp^4)$}, where the maximum value for ${\cal G}'$ in the
squeezed configuration is independent from the model parameters:
\mbox{${\cal G}'^{\rm \;sqz}_{\rm max}=\frac14$}.

\section{Conclusions}

Cosmic structure originates from the growth of quantum fluctuations during a 
period of cosmic inflation in the Early Universe. The particle production 
process generates an almost scale invariant spectrum of superhorizon 
perturbations of suitable fields, e.g. light scalar fields. These perturbations
give rise to the primordial density/curvature perturbation via a multitude of 
mechanisms (e.g. inflaton, curvaton, inhomogeneous reheating etc.). Observables
such as the spectral index $n_s$ or the non-linearity parameter $\fnl$ will 
soon exclude whole classes of models. Indeed, the Planck satellite observations
are expected to increase precision up to \mbox{$\fnl={\cal O}(1)$}.

Recently, the possibility that cosmic vector fields contribute or even generate
the curvature perturbation $\zeta$ (e.g. through the vector curvaton 
mechanism) is 
being explored. If it is so then vector fields can produce distinct signatures
such as correlated statistical anisotropy in the spectrum and bispectrum of 
$\zeta$. WMAP 
observations allow up to 30\% statistical anisotropy in the spectrum but
the Planck satellite mission is expected to reduce this bound down to 2\%
\cite{planck}, if
statistical anisotropy is not observed. Anisotropy in the non-Gaussianity can 
be dominant, which means that $\fnl$ may feature an angular modulation on the 
microwave sky.

The above suggest that cosmological observations allow for detailed modelling 
and open a window to fundamental physics complementary to Earth based 
experiments such as the LHC.

\ack
I would like to thank my collaborators Mindaugas Kar\v{c}iauskas, 
David~H.~Lyth, Yeinzon Rodriguez-Garcia and Jacques~M.~Wagstaff.
This work was supported by the European Union through the Marie Curie Research 
and Training Network ``UniverseNet" (MRTN-CT-2006-035863).
My participation to the NEB-14 conference 
was supported by a Royal Society Conference Grant.

\end{document}